\documentstyle[12pt]{article} 
 1                                        
\def\vec#1{\mbox{\protect\boldmath $ #1 $}}                                

\begin{document}                                                              
\begin{center}                                                                
{ \large\bf GRAVITATIONAL EQUILIBRIUM in THE PRESENCE of a POSITIVE COSMOLOGICAL CONSTANT \\}
\vskip 2cm  
{ Marek Nowakowski, Juan-Carlos Sanabria and Alejandro Garcia \\}                              
Departamento de Fisica, Universidad de los Andes, A.A. 4976,\\
Santafe de Bogota, D.C., Colombia
\end{center}
\vskip .5cm

\begin{abstract} 
We reconsider the virial theorem in the presence of a positive cosmological constant
$\Lambda$. Assuming steady state, we derive an inequality of the form $\rho \geq
A(\Lambda/4\pi G_N)$ for the mean density $\rho$ of the astrophysical object. 
The parameter $A$ depends only on the shape of the object. With a minimum
at $A_{sphere}$ = 2, its value can increase by several orders of magnitude
as the shape of the object deviates from a spherically symmetric one. 
This, 
among others, indicates that flattened matter distributions like 
e.g. clusters or superclusters, 
with low density, cannot be in gravitational equilibrium.  
\end{abstract}                                                                

Due to its wide range of application as well as its generality, the virial
theorem plays an important role in astrophysics.
To derive the virial theorem, one requires only the collisionless 
Boltzmann equation. 
Assuming steady state, one of the important applications of the virial theorem is to deduce the
mean density of astrophysical objects like galaxies, clusters and superclusters by observing velocities
of a `test-body' around them. It is clear that for conglomeration of matter, spread over
large enough scales, the Hubble expansion of the universe  will, in principle, oppose the gravitational 
equilibrium. It is also known that a positive cosmological constant $\Lambda \ge 0$ accelerates this 
expansion (for a review on the cosmological constant and its problems, see \cite{weinberg}). 
If, as recent measurements seem to indicate \cite{newuniv}, a positive cosmological constant enters
the Einstein's equations, the resulting space-time in the Newtonian limit (called Newton-Hooke space
\cite{bacry}) will inherit the expansion due to the $\Lambda$-term in the form of an {\it external} force. In 
this limit of the Einstein's equations we can rederive the virial theorem and evaluate the conditions under
which a steady state for a collection of matter is reached when, 
as it is the case here, we have two opposing forces: the attractive Newtonian force and the 
repulsive external $\Lambda$-force. In exploring the astrophysical significance of $\Lambda > 0$, we will 
make use of the revised virial theorem (other approaches to eventual astrophysical
effects of $\Lambda$ have been discussed in \cite{me} and \cite{others}). 

Before going into medias res of the virial theorem, we mention first some salient features of the
Newtonian limit itself, including now the cosmological constant \cite{me}. This limit is, 
in many respects, quite different from the $\Lambda=0$ case. The Newtonian limit of general relativity
is given in the form of a Poisson equation for the gravitational potential $\Phi$ which has to satisfy  
\begin{equation} \label{NL}
\vert \Phi \vert \ll 1
\end{equation}
In our case the Poisson equation reads
\begin{equation} \label{poisson}
\triangle \Phi = (4 \pi G_N) \rho -\Lambda
\end{equation}
with $G_N$ being the Newton's constant. Trivially, to solve (\ref{poisson}) we need some boundary conditions. 
In case of $\Lambda =0$, the latter is taken to be of Dirichlet type and is put at an infinite distance
$R$ i.e. $\Phi\vert_{R=\infty}={\rm const}$. This is consistent since any Newtonian potential will fall off
at large distances as $1/r$. For $\Lambda \neq 0$ there is, besides the equation (\ref{poisson}), yet another 
source of information about the Newtonian limit. This is the Schwarzschild solution for $g_{\mu \nu}$ of 
spherically symmetric bodies. Via $g_{00} = -(1 +2\Phi)$, we can then infer from this solution the potential
of a point-like and spherically symmetric body. It reads
\begin{equation} \label{pointlike}
\Phi (r)= -{G_N M \over r} - { 1 \over 6}\Lambda r^2
\end{equation}
Combining this with equation (\ref{NL}), we obtain two constraints of the validity region of the
Newtonian limit, namely
\begin{equation} \label{maxmass}
{\cal M}_{max}(\Lambda) \equiv {2\sqrt{2} \over 3}{1 \over G_N\sqrt{\Lambda}} \gg M
\end{equation}
and
\begin{equation} \label{minmax2}
{\cal R}_{max}\simeq \sqrt{6 \over \Lambda}\left[1 - {1 \over 3\sqrt{3}}{M 
\over{\cal M}_{max}}
\right ] \gg r \gg{\cal R}_{min}\simeq G_NM\left[1 -{1 \over 54}\left({M 
\over{\cal M}_{max}}\right)^2\right]
\end{equation}
Note that in case of $\Lambda =0$ neither $M_{max}$ nor $R_{max}$ exist. On account of $R_{max} \gg r$, 
we cannot anymore put the Dirichlet boundary condition at infinity if we want to have a consistent 
Newtonian limit. We have to rather choose some {\it finite} distance $R$ where we can set $\Phi\vert_R=0$.
The solution of the Poisson equation (\ref{poisson}) with this boundary condition can then be written as
\begin{eqnarray} \label{solution2}
\Phi(\vec{x}) =&-&G_N \int_{V'}d\vec{x'}{\rho(\vec{x'}) \over \vert
\vec{x} -\vec{x'}\vert}
- {1 \over 6} \Lambda \vert \vec{x}\vert^2 \nonumber \\
&+& G_N\int_{V'}d\vec{x'}G'(\vec{x}, \vec{x'})\rho(\vec{x'}) 
\end{eqnarray}
with 
\begin{equation} \label{G'}
G'(\vec{x}, \vec{x'}) \equiv 4\pi  
\sum_{l=0}^{\infty}\sum_{m=-l}^{l}{Y^{\ast}_{lm}(\theta', \varphi')
Y_{lm}(\theta, \varphi) \over 2l+1} 
{r^l_< r_>^l \over R^{2l+1}}
\end{equation}
and
$r_< = {\rm min}(\vert \vec{x} \vert , \vert \vec{x'} \vert)$,
$r_> = {\rm max}(\vert \vec{x} \vert , \vert \vec{x'} \vert)$. Hence, there are now two effects of 
the $\Lambda$-term in the potential: a direct term in the form of an external potential $-(1/6)\Lambda
\vert \vec{x}\vert^2$ and an indirect one, involving $G'(\vec{x}, \vec{x'})$, which appears because we 
were forced to choose the boundary condition at a finite distance. 
If $R$ is some fraction of $R_{max}$ (the only distinguished large radius available), but still quite 
sizeable, the indirect contribution to $\Phi$ will be suppressed by powers $R^{-n}$ and, in the first
approximation we can neglect its appearance in the virial theorem.

Having discussed the Newtonian limit to some extent, we derive next the virial theorem including the 
external $-(1/6)\Lambda \vert \vec{x} \vert^2$ potential. We could start with 
the collisionless Boltzmann 
equation, but for simplicity we will follow here a more pedestrian approach by considering a discrete
collection of massive objects enumerated by $\alpha, \beta =1,2,3,...,N$. The inertia tensor is then
given by
\begin{equation} \label{inertia}
I_{jk}=\sum_{\alpha =1}^N m_{\alpha}x_j^{\alpha}x_k^{\alpha}
\end{equation}
Differentiating this tensor twice with respect to time, using Newton's equation of motion for 
${\ddot  x}_j^{\alpha}$ and defining the kinetic energy tensor
$K_{jk}$ and the corresponding potential energy tensor $W_{jk}$ by
\begin{eqnarray} \label{tensor}
K_{jk}&=& {1 \over 2}\sum_{\alpha =1}^N m_{\alpha} 
{\dot x}^{\alpha}_j
{\dot x}_k^{\alpha}
\nonumber \\
W_{jk}&=&W_{kj}=-{1 \over 2}G_N\sum_{\alpha \neq \beta =1}^N m_{\alpha}
m_{\beta} {(x_j^{\alpha}-x_j^{\beta})(x_k^{\alpha}-x_k^{\beta}) \over \vert \vec{x}^{\alpha}-
\vec{x}^{\beta}\vert^3}
\end{eqnarray}
we readily arrive at
\begin{equation} \label{virial}
{d^2 I_{jk} \over dt^2}=4K_{jk}+2W_{jk} +{2 \over 3}\Lambda I_{jk}
\end{equation}
which is the virial theorem. It is worthwhile stressing that given $K_{jk}$ and $W_{jk}$ the presence of $\Lambda$
makes the virial theorem a differential equation for $I_{jk}$. However, lacking a priori this information, 
one can proceed as follows. First we take the trace of equation (\ref{virial}) and define 
$K \equiv {\rm Tr}(K_{jk})$, $W \equiv {\rm Tr}(W_{jk})$, $I \equiv {\rm Tr}(I_{jk})$. Next we assume a steady
state i.e. ${d^2 I \over dt^2}=0$. Then equation (\ref{virial}) becomes
\begin{equation} \label{virial2}
2K+W+{1 \over 3}\Lambda I=0
\end{equation}
Had we started with the collisionless Boltzmann equation, the final result would have been the same. 
It is now convenient to 
pass to continuous systems for which
\begin{eqnarray} \label{continous}
I&=&\int_V d\vec{x} \rho(\vec{x})\vert \vec{x}\vert^2 \nonumber \\
K&=& {1 \over 2} \int_Vd\vec{x} \rho(\vec{x}) \bar{v}^2 \nonumber \\
W&=& {1 \over 2}\int_V d\vec{x} \rho(\vec{x}) \Phi_N(\vec{x})
\end{eqnarray}
where $\bar{v}$ is the averaged velocity and $\Phi_N(\vec{x})$ is the Newtonian part of the solution in 
(\ref{solution2})
(proportional to $G_N$ neglecting the third term involving $G'(\vec{x}, \vec{x'})$). Since $\Phi_N \le 0$, it 
follows that
$W \le 0$. On the other hand K is positive definite. This leads to the basic inequality of gravitational equilibrium 
with a non-zero $\Lambda$
\begin{equation} \label{inequality}
-{1 \over 3} \Lambda I + \vert W \vert \geq 0
\end{equation}
Note that this inequality is trivially satisfied if $\Lambda =0$. Any astrophysical object which does not obey this
inequality, cannot be in a gravitational equilibrium. Clearly, equation (\ref{inequality}) is a relation between the 
density of the body and its shape (geometry), the latter encoded in the integrals over the volume of the body. 
One way to
exploit (\ref{inequality}) would be to model the density for the object under consideration (as it is often done
in astrophysics) by assuming a functional form $\rho=\rho(\vec{x}; \lambda_i)$ where the $\lambda_i$'s are some parameters
(often of dimension of mass or length). Then the inequality (\ref{inequality}) would convert into a relation among these parameters. It 
is, however, more transparent to work with a constant mean density
$\rho$. Defining then
\begin{equation} \label{W'}
I=\rho I', \,\,\,\, W=-{1 \over 2} \rho^2 G_N W'
\end{equation}
we can recast the inequality (\ref{inequality}) into the form
\begin{eqnarray} \label{theineq}
\rho &\geq& A\rho_{\rm vac}\nonumber \\
A &\equiv& {16 \pi \over 3} {I' \over W'}, \,\,\,\, \rho_{\rm vac} \equiv {\Lambda \over 8 \pi G_N}
\end{eqnarray}
To search for the effects of $\lambda$, inequality (\ref{theineq}) has an appealing form as it is solely expressed 
through the mean density and a parameter $A$ which depends only on the geometry. 
It is obvious that the larger the value of 
 $A$ is, the more will be the importance which we can attribute to $\Lambda$ with regard to gravitational equilibrium.

To investigate the possible sizes of $A$ we consider the volume of the body bounded by smooth surfaces of second
degree. With the triaxial ellipsoid many different volumes can then be approximately modeled. For the sphere
of radius $R_0$ we obtain $W'=(32/15)\pi^2R_0^5$ and $I'=(4/5)\pi R_0^5$ resulting into
\begin{equation} \label{sphere}
A_{\rm sphere}=2
\end{equation}
As we will see below, this seems to be the smallest possible value of $A$. For an ellipsoid with three axes
$a$, $b$ and $c$ the integrals give $W'=(16/15)\pi^2 abc Ja^2$ and $I'=(4/15)\pi abc(a^2 +b^2 +c^2)$ where the 
parameter $J$ depends on the type of the ellipsoid i.e. on the remnant rotational symmetry. The parameter $A$
for an ellipsoid has the general form
\begin{equation} \label{ellipsoid}
A_{\rm ellipsoid}={4 \over 3}{a^2 +b^2 +c^2 \over Ja^2}
\end{equation}   
We quote now the parameters $J$'s for the three different types of ellipsoids \cite{J}
\begin{eqnarray} \label{J}
J_{\rm prolate}&=& {1-e^2 \over e} \ln\left({1+e \over 1-e}\right), \,\,\, a=b < c \nonumber \\
e &\equiv& \sqrt{1 -\left({a \over c}\right)^2} \nonumber \\
J_{\rm oblate}&=& 2{\sqrt{1-e'^2} \over e'} \sin^{-1}e', \,\,\,\, a=b > c \nonumber \\
e'&\equiv& \sqrt{1- \left({c \over a}\right)^2} \nonumber \\
J_{\rm triaxial}&=& 2{bc \over a^2} {F(\phi, k) \over \sin\phi}, \,\,\, a>b>c \nonumber \\
\phi &\equiv& \cos^{-1}{c \over a}, \,\,\,\, k\equiv \sqrt{{1-({b\over a})^2 \over 1-({c \over a})^2}}
\end{eqnarray}
where $F(\phi, k)$ is the elliptic integral of the first kind. The corresponding three different 
parameters for $A$ can be calculated to be
\begin{eqnarray} \label{A}
A_{\rm prolate}= &{4\over 3}&\left({c\over a}\right)^4\left( 1+2\left({a \over c}\right)^2\right)
{e \over \ln\left(
{1+e \over 1-e}\right)} \nonumber \\
\buildrel c \gg a \over \longrightarrow &{2 \over 3}&\left({c \over a}\right)^4
{1 \over \ln{c \over a}} 
\nonumber \\
A_{\rm oblate}=&{4 \over 3}& e'\left({a \over c}\right)\left( 1 + \left({c \over \sqrt{2} a}\right)^2
\right) {1 \over \sin^{-1}e'} \nonumber \\
\buildrel a \gg c \over \longrightarrow &{8 \over 3\pi }& \left({a \over c}\right) \nonumber \\
A_{\rm triaxial}= &{2 \over 3}&{a^2 +b^2+c^2 \over bc}{\sin\phi \over F(\phi, k)} \nonumber \\
\buildrel a \gg b > c \over \longrightarrow &{2 \over 3}&\left({a \over b}\right)\left({a \over c}\right){1 \over 
\ln4\sqrt{{a^2 \over b^2 -c^2}}}
\end{eqnarray}
To appreciate the final results (eq.(\ref{theineq}) together with (\ref{A})), let us take an example of a flattened
prolate ellipsoid with $(c/a)=10 (10^2)$. The resulting $A_{\rm prolate}$ is then $10^3 (10^7)$. 
Given first the 
fact that the nowadays preferred  value of $\rho_{\rm vac}$ is $(0.7-0.8)\rho_{\rm crit}$ ($\rho_{\rm crit} =
1.789 \times 10^{-29} h_0^2 {\rm gcm}^{-3}$, $0.6 < h_0 < 0.8$) and secondly that in comparison to 
$\rho_{\rm vac}$ the density of a typical elliptic galaxy is $(0.1-2) \times 10^{-26} {\rm gcm}^{-3}$, 
we can 
safely state that inequality (\ref{theineq}), and hence the effect of $\Lambda$ in considering gravitational 
equilibrium, is of astrophysical relevance. 
The type of astrophysical objects (galaxies, clusters and 
superclusters) we are investigating may or may not be in gravitational 
equilibrium. In both cases, the above results
seem relevant. In the first case, with a fixed density $\rho$, we would expect that the deviation of the 
rotational symmetry is indeed small. In trying to reach the equilibrium, the objects will reduce the ratio of
their axes. This is a direct consequence of the inequality (\ref{theineq}) and the values of $A$ given in
(\ref{sphere}) and (\ref{A}). As such, this is also a consequence of $\Lambda > 0$. If, in contrast the 
inequality (\ref{theineq}) is not satisfied, the gravitational steady state is not reached and we should be 
careful in extracting the densities from the virial theorem in such situations. First, for low density objects
the effect of $\Lambda$ should be included in the virial theorem (eq. (\ref{virial2})). Furthermore, 
the inequality (\ref{theineq}) can always serve as a check if the assumed equilibrium indeed holds.

Let us now consider objects which are most likely not in a steady state. These are superclusters which appear 
mostly in a flattened form and have low densities. Even for richly populated superclusters with $\rho \sim
10^{-29} {\rm gcm}^{-3}$, their deviation from spherical symmetry would result in a violation of the inequality
(\ref{theineq}). The situation worsens with superclusters of densities $(10^{-20}-10^{-31}) {\rm gcm}^{-3}$
which are known to exist \cite{super2}.

For clusters the answer to the question whether they are in gravitational equilibrium is less certain. With
a typical cluster density of $10^{-18} {\rm gcm}^{-3}$, the crucial ingredient now is how much their shapes 
deviate from spherical symmetry. We speculate that low density clusters with large off-sphericity could exist
and a systematic survey (which goes beyond the scope of this letter) 
would be welcome. If most  
clusters have a geometry of 
low eccentricities and sufficiently large densities, the line of argument is similar to the case of elliptic
galaxies discussed below.

Elliptic galaxies  with highest eccentricities are known to be of the type E7. This corresponds to $(c/a)=3.3$. Why do 
we not see galaxies with a higher ratio than that corresponding to E7? Without invoking galactic dynamics, 
the answer could be the inequality (\ref{theineq}). In view of our results 
objects trying to reach the equilibrium, can do that, given enough time, by lowering the ratio of their axes. 
Apparently, galaxies have done exactly this. It should not be forgotten that our results include a positive
cosmological constant $\Lambda$. If $\Lambda =0$, from the point of view of the virial theorem, galaxies with a 
high ratio $(c/a)$ could, in principle, be also in gravitational equilibrium.

The significance of all results, presented in the paper, will grow with
increasing value of $\Lambda$. In 
so-called quintessence models $\Lambda$ is effectively a parameter which depends on the epoch 
\cite{quintessence}. 
As such it could have 
been larger in the past. The investigation of structure formation in such models could be 
of relevance.

Two issues remain to be explored in more detail in the future. Looking back at equation (\ref{virial})
and assuming non-equilibrium, additional insight could be gained if we could solve the differential
equation for $I_{jk}(t)$. This requires that we succeed in obtaining more information on
$K_{jk}(t)$ and $W_{jk}(t)$ in terms of new differential equations or otherwise. Secondly, 
the expansion of the universe which can, in principle, oppose the gravitational steady state is 
in our equations only due to the $\Lambda$-term. It would be certainly worthwhile to make an attempt
to include the full expansion. We expect that this would strengthen our results.

\end{document}